\documentclass[fleqn, usenatbib]{mnras}

\usepackage{times}
\usepackage[utf8]{inputenc}
\usepackage[T1]{fontenc}
\usepackage{ae,aecompl}

\usepackage{graphicx} 
\usepackage{amsmath}  
\usepackage{amssymb}  

\newcommand{\be}{\begin{equation}}
\newcommand{\ee}{\end{equation}}
\newcommand{\beq}{\begin{eqnarray}}
\newcommand{\eeq}{\end{eqnarray}}
\newcommand{\ba}{\begin{align}}
\newcommand{\ea}{\end{align}}

\def\f{\frac}

\title[Low-mass neutron stars]{Low-mass neutron stars: universal
  relations, the nuclear symmetry energy and gravitational radiation}

\author[Hector O. Silva et al]{
Hector O. Silva$^{1}$\thanks{E-mail: hosilva@phy.olemiss.edu},
Hajime Sotani$^{2}$\thanks{E-mail: sotani@yukawa.kyoto-u.ac.jp},
Emanuele Berti$^{1,3}$\thanks{E-mail: eberti@olemiss.edu}
\\
$^{1}$Department of Physics and Astronomy, The University of Mississippi, University, Mississippi 38677, USA\\
$^{2}$Division of Theoretical Astronomy, National Astronomical Observatory of Japan, 2-21-1 Osawa, Mitaka, Tokyo 181-8588, Japan\\
$^{3}$CENTRA, Departamento de F\'isica, Instituto Superior T\'ecnico, Universidade de Lisboa, Avenida Rovisco Pais 1, 1049 Lisboa, Portugal
}

\date{Accepted XXX. Received YYY; in original form ZZZ}

\pubyear{2016}

\begin{document}
\label{firstpage}
\pagerange{\pageref{firstpage}--\pageref{lastpage}}
\maketitle

\begin{abstract}
The lowest neutron star masses currently measured are in the range
$1.0-1.1~M_\odot$, but these measurement have either large
uncertainties or refer to isolated neutron stars. The recent claim
of a precisely measured mass $M/M_{\odot} = 1.174 \pm 0.004$
\citep{Martinez:2015mya} in a double neutron star system suggests
that low-mass neutron stars may be an interesting target for
gravitational-wave detectors.  Furthermore, \cite{Sotani:2013dga}
recently found empirical formulas relating the mass and surface
redshift of nonrotating neutron stars to the star's central density
and to the parameter $\eta\equiv (K_0 L^2)^{1/3}$, where $K_0$ is
the incompressibility of symmetric nuclear matter and $L$ is the
slope of the symmetry energy at saturation density.  Motivated by
these considerations, we extend the work by \cite{Sotani:2013dga} to
slowly rotating and tidally deformed neutron stars. We compute the
moment of inertia, quadrupole moment, quadrupole ellipticity, tidal
and rotational Love number and apsidal constant of slowly rotating
neutron stars by integrating the Hartle-Thorne equations at second
order in rotation, and we fit all of these quantities as functions
of $\eta$ and of the central density. These fits may be used to
constrain $\eta$, either via observations of binary pulsars in the
electromagnetic spectrum, or via near-future observations of
inspiralling compact binaries in the gravitational-wave spectrum.
\end{abstract}



\section{Introduction}
\label{sec:intro}

The equilibrium of spherically symmetric, nonrotating neutron stars
(NSs) in general relativity is governed by the
Tolman-Oppenheimer-Volkoff (TOV) equations, that follow from
Einstein's equations with a perfect-fluid stress energy
tensor~\citep[see e.g.][]{Stergioulas:2003yp,Shapiro:1983du,FriedmanStergioulas}.
When supplemented with an equation of state (EOS) relating the density and
pressure of the perfect fluid, the TOV equations form a closed system
of ordinary differential equations, whose solutions are obtained (in
general) by numerical integration.  The solutions form a
single-parameter family, where the parameter can be chosen to be the
central total energy density $\rho_{\rm c}$.
Despite recent progress, the EOS is still largely unknown at the
energy densities $\rho>\rho_0$ (where
$\rho_0/c^2 \equiv 2.68 \times 10^{14}$ g/cm$^3$ is the nuclear
saturation density) characterizing the NS core.  Uncertainties in the
EOS translate into uncertainties in the NS mass-radius relation: for a
typical NS mass $M\sim 1.4 M_\odot$, EOSs compatible with our current
knowledge of nuclear physics predict radii $R$ ranging between $6$ and
$16$~km~\citep{Steiner:2012xt}.

Unlike black holes, which are vacuum solutions of Einstein's
equations, NS structure depends on the coupling of gravity with
matter. Therefore NSs can probe (and possibly rule out) theories of
gravity that are close to general relativity in vacuum, but differ in
the description of the coupling between matter and gravity in the
strong-field regime \citep{Berti:2015itd}.  In fact, given the
strength of their gravitational field, the high density of matter at
their cores and the existence of pulsars with fast spin and large
magnetic fields, NSs are ideal laboratories to study all fundamental
interactions~\citep{Lattimer:2004pg,Lattimer:2006xb,Lattimer:2015nhk,Psaltis:2008bb}.
However, tests of strong gravity with NSs are made more difficult by
two fundamental degeneracies: (i) uncertainties in the EOS can mimic
the modifications to the bulk properties of NSs that may be induced by
hypothetical strong-field modifications of general relativity; (ii)
different theories of gravity can give rise to similar modifications
in the bulk properties of NSs~\citep{Glampedakis:2015sua}.

These issues are partially alleviated by recently discovered
``universal'' (EOS-independent) relations showing that rotating NSs
are, in fact, relatively simple objects. Let $M$ be the mass of a
nonrotating star, $J$ the angular momentum, $\chi=J/M^2$ the
dimensionless spin, $I$ the moment of inertia, $Q$ the quadrupole
moment and $\lambda^{\rm (tid)}$ the tidal Love number, a measure of
stellar deformability (here and below we use geometrical units,
$G=c=1$).  Working in the slow-rotation approximation,
\cite{Yagi:2013awa,Yagi:2013bca} discovered that universal
(EOS-independent) ``$I$-Love-$Q$'' relations connect the three
normalized quantities $\bar I=I/M^3$,
$\bar \lambda^{\rm (tid)}=\lambda^{\rm (tid)}/M^5$ and
$\bar Q= - Q^{\rm (rot) \ast}/(M^3\chi^2)$. Subsequent work relaxed
the slow-rotation approximation, showing that the universality still
holds~\citep{Doneva:2013rha,Pappas:2013naa,Chakrabarti:2013tca,Yagi:2014bxa}.

Most investigations of relativistic stellar structure focus on NSs
with the ``canonical'' $1.4 M_{\odot}$ mass or higher. From a nuclear
physics standpoint, recent measurements of masses
$M\gtrsim 2M_{\odot}$ have ruled out EOS models that are unable to
support such high masses
\citep{Demorest:2010bx,Antoniadis:2013pzd}\footnote{Here we will
  consider some EOS models that do not respect this constraint. This
  is because we are primarily interested in densities
  $\rho\sim \rho_0$, and (conservatively) we assume no correlation
  between the EOS near the saturation density and the EOS at higher
  densities \citep[cf.][]{Steiner:2014pda}.}.
Large-mass NSs are more compact, and therefore more interesting for
tests of strong gravity. From an astrophysical point of view, the
large-mass regime is also interesting to improve our understanding of
core-collapse physics. Observations of NS binaries (particularly via
radio pulsars) and black hole X-ray binaries indicate that there may
be a mass gap between the two populations: the highest measured NS
masses just exceed $2M_{\odot}$~\citep{Lattimer:2012nd}, while black
hole masses may only start at $\sim 4$--$5.5M_{\odot}$
\citep{Ozel:2010su,Farr:2010tu}, depending on the assumed shape of the
distribution \citep[but see][who point out that selection biases could
yield lower black hole masses]{Kreidberg:2012ud}.  Gravitational-wave
observations of merging compact binaries will offer a unique
opportunity to probe the existence of this ``mass gap''
\citep{Dominik:2014yma,Mandel:2015spa,Littenberg:2015tpa,Stevenson:2015bqa,Belczynski:2015tba,Chatziioannou:2014coa}.

Our focus here is instead on low-mass NSs. There are observational and
theoretical reasons why this regime is interesting.
The lowest well-constrained NS masses currently measured are in the
range $1.0$--$1.1~M_\odot$ \citep{Lattimer:2012nd}. Recently
\cite{Martinez:2015mya} claimed a precise measurement of
$M/M_{\odot} = 1.174 \pm 0.004$ in a double NS system with large mass
asymmetry.
While the minimum mass of a star constructed from a cold dense matter
EOS is quite small ($< 0.1M_{\odot}$), the minimum mass of a hot
protoneutron star is considerably larger, in the range
$0.89$--$1.13M_{\odot}$ for the models considered by
\cite{Strobel:1999vn}.
This minimum mass provides a practical lower bound on NS masses formed
from supernovae, unless lower-mass stars form by fragmentation
\citep[see e.g. the speculative scenario
of][]{Popov:2006ki}. Estimates based on the baryonic mass of the iron
core of the supernova progenitor give a minimum mass of
$\sim 1.15$--$1.2M_{\odot}$, as discussed in Sec.~3.3 of
\cite{Lattimer:2012nd}. \cite{Tauris:2015xra} estimate that the
minimum mass of a NS formed in an ultra-stripped supernova is
$1.1M_{\odot}$. Note that uncertainties in supernova physics affect
all of these bounds, and (if confirmed) the recent observations of
\cite{Martinez:2015mya} are only marginally compatible with the iron
core bound.
To summarize: it is commonly believed that the minimum mass of NSs in
the universe should be around the minimum observed mass
($\sim 1M_{\odot}$) and that NS masses $\lesssim 1.2 M_{\odot}$ would
challenge the paradigm of NS formation by gravitational collapse, but
these conclusions are uncertain due to our limited understanding of
supernova physics. Therefore the discovery of low-mass NSs may give us
important clues on their formation mechanism: for example,
observations of NSs with mass $M\lesssim 1M_{\odot}$ could validate
the astrophysical viability of the proto-NS fragmentation scenario
proposed by \cite{Popov:2006ki}.

Another key motivation for this work is that the low-mass regime is
sensitive to -- and carries information on -- the isospin dependence
of nuclear forces, and in particular on the nuclear symmetry energy
\citep{Steiner:2004fi,Tsang:2012se,Li:2013ola,
Lattimer:2014sga,Li:2014oda,Newton:2015xza}.
\cite{Sotani:2013dga} recently computed the structure of low-mass
nonrotating NSs for a wide range of EOSs. They found that their mass
$M$ and surface redshift $z$ can be fitted by simple functions of the
central density $\rho_{\rm c}$ and of the dimensionful parameter
\be
\label{def:eta}
\eta\equiv \left(K_0 L^2\right)^{1/3} \,,
\ee
where $K_0$ is the incompressibility of symmetric nuclear matter and
$L$ is the slope of the symmetry energy at saturation density (note
that $K_0$, $L$ and $\eta$ all have units of energy).
Therefore, at least in principle, measurements of $M$ and $z$ could be
used to constrain $\eta$; in fact, the NS radius is highly correlated
with the NS matter pressure at densities close to nuclear saturation
density.  A practical complication is that the determination of $z$
and of the stellar radius, e.g. via photospheric radius expansion
bursts and thermal emissions from quiescent low-mass X-ray binaries,
is model-dependent and affected by systematic errors. Therefore, at
present, no individual observation can determine NS radii to better
than $\sim 20\%$ accuracy. This translates into nearly a $100\%$ error
in the determination of $L$, since $L\sim R^4$
\citep{Lattimer:2014sga}.

A possible way to circumvent this problem is to rely on the fact that
all NSs in nature are spinning.
Considering rotating NSs is of interest because near-future
experiments in the electromagnetic spectrum -- such as
NICER~\citep{2012SPIE.8443E..13G}, LOFT~\citep{Feroci:2012qh},
Astro-H~\citep{Takahashi:2012jn} and SKA~\citep{Watts:2014tja} -- or
in the gravitational-wave spectrum -- such as Advanced LIGO~\citep{TheLIGOScientific:2014jea}, Advanced
Virgo~\citep{TheVirgo:2014hva}, KAGRA~\citep{Somiya:2011np} and the
Einstein Telescope~\citep{Sathyaprakash:2012jk}
-- could measure or constrain the additional multipoles that determine
the structure of rotating NSs or other properties (such as the ``Love
numbers'') that are related to their deformability.
Spin-orbit coupling in binary pulsars may allow us to measure the
moment of inertia \citep{Damour:1988mr,Lattimer:2004nj,Bejger:2005jy,Kramer:2009zza}
and gravitational-wave observations may be used to infer the tidal
Love numbers, as well as additional information on the EOS
\citep{Mora:2003wt,Berti:2007cd,Flanagan:2007ix,Read:2009yp,Hinderer:2009ca,Vines:2011ud,Damour:2012yf,DelPozzo:2013ala,Read:2013zra,Favata:2013rwa,Yagi:2013baa,Lackey:2013axa,Chatziioannou:2015uea,Yagi:2015pkc,Dietrich:2015pxa}.
Quite remarkably, measurements of the moment of inertia within an
accuracy $\sim 10\%$ {\em alone} can yield tight constraints on the
pressure over a range of densities \citep{Steiner:2014pda}.  The
correlation between the moment of inertia and the tidal deformability
(the ``$I$-Love'' relation) is very tight for massive NSs
\citep{Yagi:2013awa,Yagi:2013bca}, but not so much in the low-mass
regime: see e.g. Fig.~\ref{fig:ILQ} below.

One of the main results of the present work is that all of the
properties of rotating and tidally deformed stars can be expressed as
simple functions of $\rho_{\rm c}$ and $\eta$. Therefore measurements
of any two bulk properties of a low-mass NS -- for example, the mass
$M$ and the moment of inertia $I$ -- can be used to determine a region
in the $\rho_c$-$\eta$ plane, permitting to estimate $\eta$.


The plan of the paper is as follows. In Section~\ref{sec:eos} we
discuss the EOS models used in this paper and the properties of
nuclear matter that are relevant in the low-mass regime. In
Section~\ref{sec:NSs} we present our numerical results for the bulk
properties of nonrotating and slowly rotating NSs, and we fit the
properties of slowly rotating NSs by nearly universal functions of the
central density $\rho_{\rm c}$ and of the parameter $\eta$. In the
concluding Section~\ref{sec:conclusions} we discuss possible
observational applications and future extensions of our work.

\section{Low-mass neutron star properties and the nuclear symmetry
  energy}
\label{sec:eos}

In this section we introduce some notation for the properties of
uniform nuclear matter near saturation density, and we describe the
EOS models used in our numerical work.

\subsection{Properties of uniform nuclear matter}

The energy of uniform nuclear matter at zero temperature can be
expanded around the saturation point of symmetric nuclear matter
(i.e., matter composed of an equal number of neutrons and protons). If
$n_b$ is the nucleon number density and $\alpha\equiv (n_n-n_p)/n_b$,
where $n_n$ ($n_p$) is the neutron (proton) number density, the bulk energy
per nucleon $w$ of uniform nuclear matter can be written as
\be
w=w_0+\f{K_0}{18n_0^2}(n_b-n_0)^2+\left[S_0+\f{L}{3n_0}(n_b-n_0)\right]\alpha^2\,,
\label{eq:snm}
\ee
where $w_0$, $n_0$ and $K_0$ are the saturation energy, the saturation
density and the incompressibility of symmetric nuclear matter, while
$S_0$ and $L$ are associated with the symmetry energy coefficient
$S(n_b)$:
\be
S_0=S(n_0)\,,  \quad L=3n_0 \left. \left(\f{dS}{dn_b}\right)
\right|_{n_b=n_0}\,.
\ee
The parameters $w_0$, $n_0$ and $S_0$ can be relatively easily
determined from empirical data for the masses and radii of stable
nuclei. The parameters $K_0$ and $L$,
which determine the stiffness of
neutron-rich nuclear matter, are more difficult to fix, and they
affect the structure of low-mass NSs.

\begin{figure}
\includegraphics[width=\columnwidth]{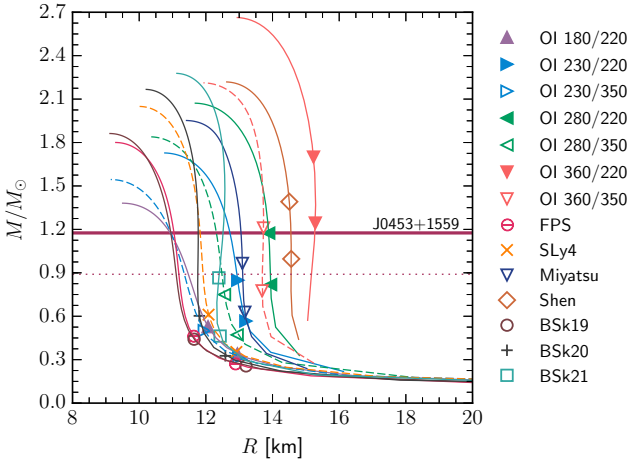}
\caption{Mass-radius relations for the EOSs adopted in this paper, and
  discussed in Sec.~\ref{sec:eos}. The curves span NS models starting
    from $u_{\rm c}=\rho_{\rm c}/\rho_0 = 0.9$ up to the value of
    $u_{\rm c}$ resulting in the maximum mass allowed by each EOS. The
  solid horizontal band corresponds to the lowest high-precision NS
  mass measurement of $M/M_{\odot} = 1.174\, \pm\, 0.004$
  \citep[from][]{Martinez:2015mya}.  The dotted horizontal line
  indicates a conservative lower bound on the mass of
  $M/M_{\odot} = 0.89$ \citep[see e.g.][]{Strobel:1999vn}.  Symbols on
  each line correspond to $u_{\rm c} = 1.5$ and $u_{\rm c} = 2.0$,
  respectively.}
\label{fig:massradius}
\end{figure}

Different EOS models are based on different theoretical and
computational approaches in nuclear physics.  In order to derive
empirical formulas expressing the properties of low-mass NSs that do
not rely on specific EOSs, following~\cite{Sotani:2013dga} we
adopt several tabulated EOS models that can be separated into three
categories:
\begin{enumerate}
\item[(1)] The phenomenological EOS model constructed by Oyamatsu and
  Iida~\citep{Oyamatsu:2002mv}. The bulk energy $w(n_b,\alpha)$ is
  constructed to reproduce Eq.~(\ref{eq:snm}) in the limit where
  $n_b\to n_0$ and $\alpha\to 0$, and the optimal values of $w_0$,
  $n_0$, and $S_0$ are determined by requiring that the density
  profile of stable nuclei (determined within the extended
  Thomas-Fermi theory for given values of $L$ and $K_0$) reproduce
  experimental nuclear data. The EOSs in this category will be labeled
  as OI $K$/$Y$, where $K = K_0$ and $Y = -K_0 S_0/(3 n_0 L)$.  At
  variance with \cite{Sotani:2013dga} we will omit the OI 180/350
  EOS, because the associated values of $K_0$ and $L$ are ruled out by
  current nuclear physics constraints (cf. Sec.~\ref{sec:expcons}).
\item[(2)] Two EOS models based on the relativistic framework. One of
  these models (Shen) is constructed within relativistic mean field theory
  together with the TM1 nuclear interaction~\citep{Shen:1998gq}; the
  second (Miyatsu) is based on the relativistic Hartree-Fock theory with the
  chiral quark-meson coupling model \citep{Miyatsu:2013hea}. The
  spherical nuclei in the crust region are determined using the
  Thomas-Fermi theory.
\item[(3)] Five EOS models based on the Skyrme-type effective
  interactions: FPS, SLy4, BSk19, BSk20, and
  BSk21~\citep{Lorenz:1992zz,Douchin:2001sv,Goriely:2010bm,Pearson:2011zz,Pearson:2012hz,Potekhin:2013qqa}.
\end{enumerate}
All of these models are {\it unified} EOSs, i.e., both the crust and
core regions can be described with the same EOS with specific values
of $K_0$ and $L$.  From the EOS tables we can compute $K_0$, $L$ and
the auxiliary dimensionful parameter $\eta$ introduced in
Eq.~(\ref{def:eta}) above, with the results listed in
Table~\ref{tab:eosparam}.
The mass-radius relations predicted by these EOS models for
nonrotating low-mass NSs are shown in Fig.~\ref{fig:massradius}.

\begin{table}
\caption{Parameters of the EOSs used in this work. EOSs which
result in NSs with a maximum mass larger than $2 M_{\odot}$
are indicated by an asterisk.}
\label{tab:eosparam}
\begin{tabular}{c c c c c}

\hline
\hline
EOS  &  $K_0$ (MeV)  &  $L$ (MeV)  &  $\eta$ (MeV)\\
\hline

OI 180/220  &  180  &  52.2  &  78.9  \\

OI 230/350  &  230  &  42.6  &  74.7  \\
OI 230/220  &  230  &  73.4  &  107  \\

OI 280/350  &  280  &  54.9  &  94.5  \\
OI 280/220$^{\ast}$  &  280  &  97.5  &  139  \\

OI 360/350$^{\ast}$  &  360  &  76.4  &  128  \\
OI 360/220$^{\ast}$  &  360  &  146  &  197  \\

\\

Shen$^{\ast}$  &  281  &  114  &  154  \\
Miyatsu  &  274  &  77.1  &  118  \\

\\

FPS  &  261  &  34.9  &  68.2  & \\
SLy4$^{\ast}$  &  230  &  45.9  &  78.5  & \\
BSk19  &  237  &  31.9  &  62.3  & \\
BSk20$^{\ast}$  &  241  &  37.9  &  69.6  & \\
BSk21$^{\ast}$  &  246  &  46.6  &  81.1  & \\
\hline
\hline
\end{tabular}
\end{table}

\begin{figure*}
 \includegraphics[width=\columnwidth]{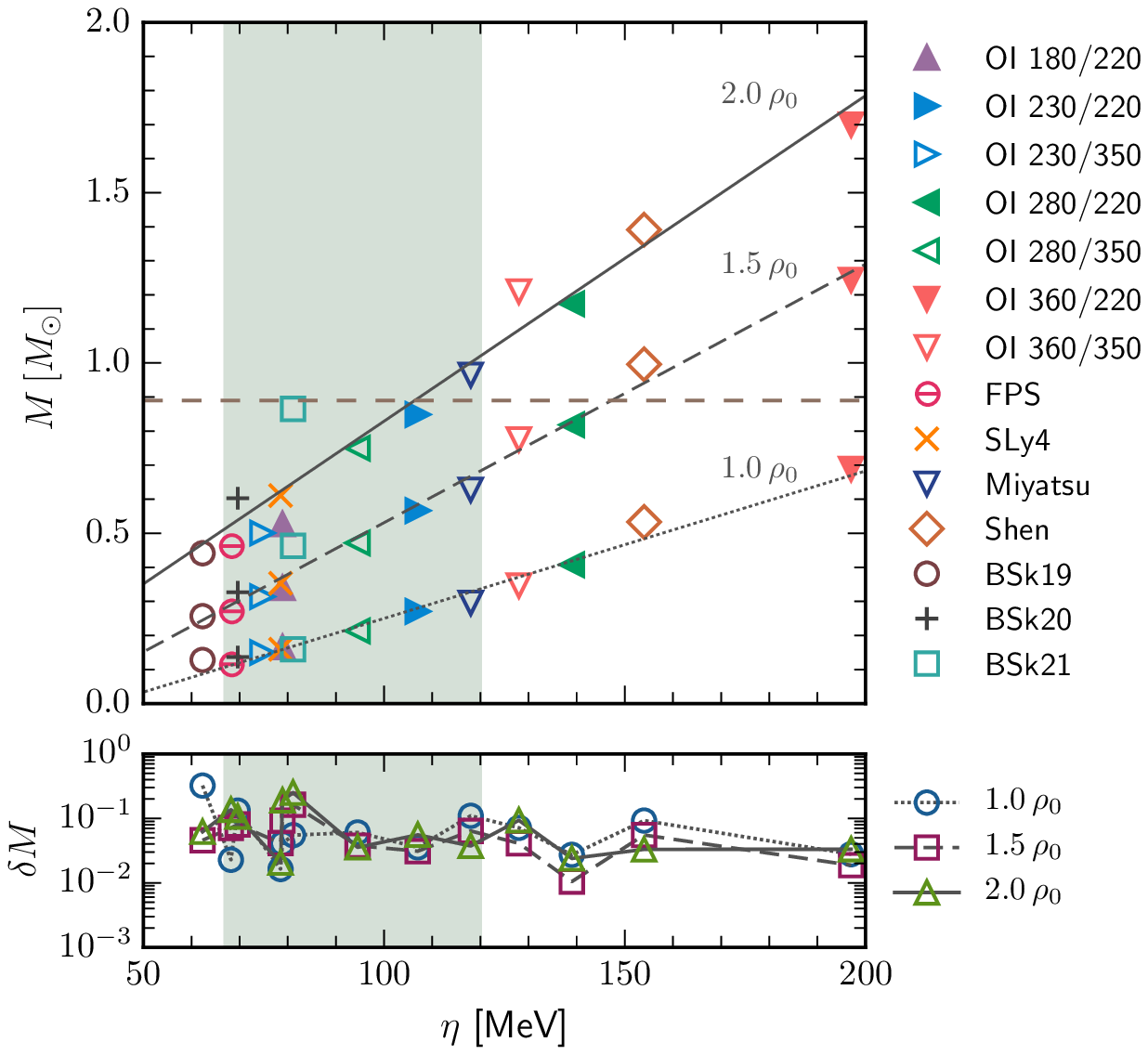}
 \includegraphics[width=\columnwidth]{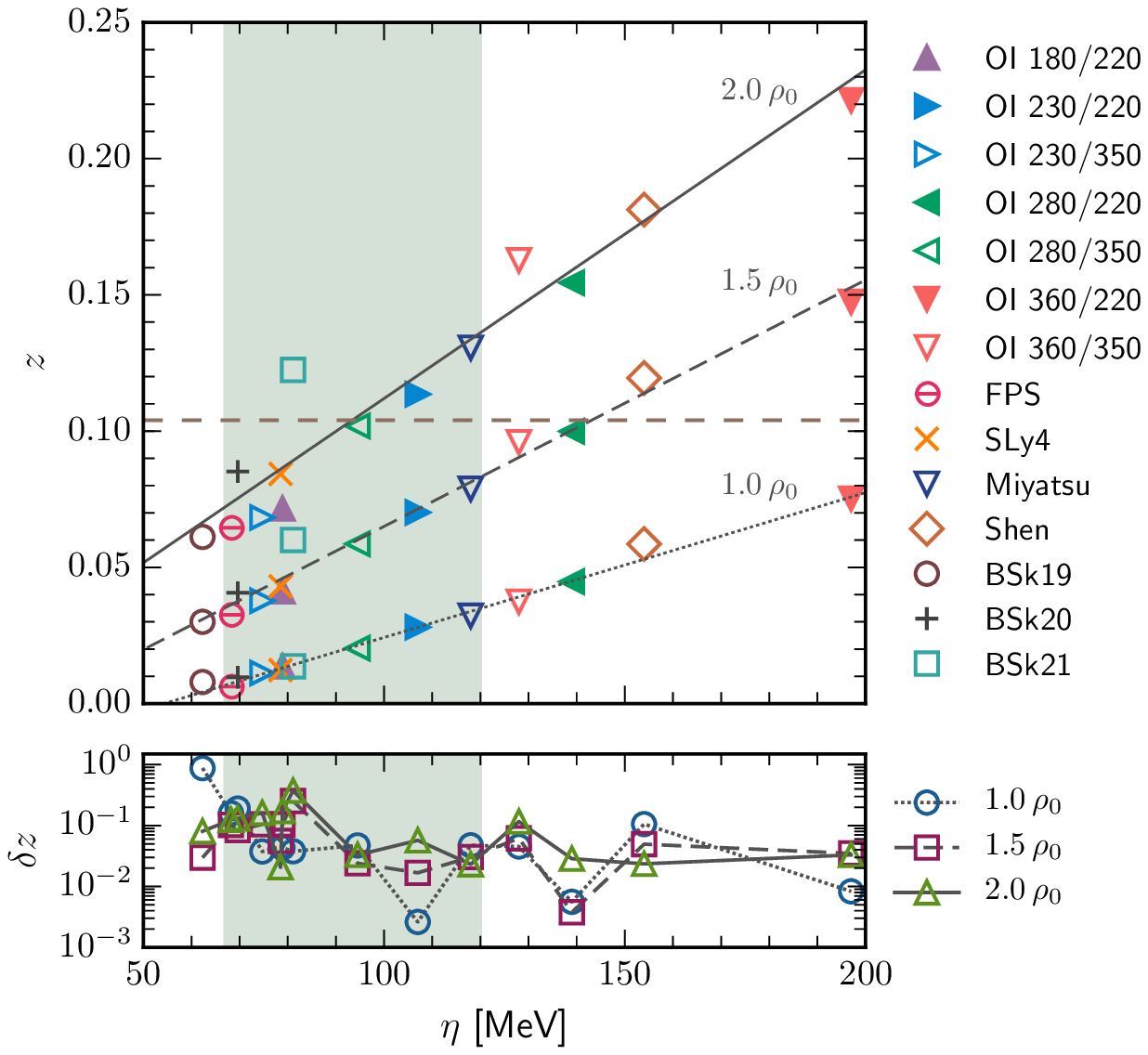}
 \caption{{\it Left}: Dependence of the NS mass $M$ for a nonrotating
   star on the parameter $\eta$ introduced in Eq.~(\ref{def:eta}) at
   three given values of the central density:
   $u_{\rm c} \equiv \rho_{\rm c} / \rho_0=1.0,\, 1.5,\, 2.0$.  {\it
     Right}: same, but for the surface gravitational redshift $z$
   defined in Eq.~(\ref{redshift}). In both cases, the solid lines
   represent the fit given in Eq.~(\ref{eq:fit_1}) using the fitting
   parameters listed in Table~\ref{tab:fitcoef}.  The lower panels
   show the relative error of the fit with respect to the numerical
   data, $|y_{\rm data} -y_{\rm fit}|/y_{\rm data}$, as a function of
   $\eta$. The shaded area corresponds to the most plausible range of
   values for $\eta$, namely $67<\eta<120$ (see
   Sec.~\ref{sec:expcons}).  This plot reproduces and extends Fig.~2
   of \citet{Sotani:2013dga}.  Horizontal dashed lines correspond to a
   NS with $M/M_{\odot} = 0.89$ (the value of $z$ was computed using
   the Shen EOS).}
 \label{fig:fits_M}
\end{figure*}

\subsection{Experimental constraints}
\label{sec:expcons}

An extensive discussion of theoretical and experimental constraints on
the properties of uniform nuclear matter can be found in
\cite{Li:2013ola}, \cite{Lattimer:2014sga} and
\cite{Newton:2015xza}. Generally accepted values of $K_0$ are in the
range $K_0 = 230 \pm 40$~MeV \citep{Khan:2013mga}. The current
consensus in the nuclear physics community is that values of $L$ in
the range $L=60 \pm 20$~MeV are plausible\footnote{\cite{Lattimer:2014sga}
  suggest a tighter plausible range of $44$~MeV$<L<66$~MeV (see the
  discussion of their Figure~1), but we will follow
  \cite{Newton:2015xza} in an attempt to be more conservative.} but
higher values are not excluded \citep{Newton:2015xza}.

There are proposals to constrain nuclear saturation parameters, and in
particular the value of $L$, via astronomical observations. This
approach is completely different from constraints based on nuclear
physics experiments. Under the assumption that the observed
frequencies of quasiperiodic oscillations (QPOs) in NSs are related to
crustal torsional oscillations, \cite{Sotani:2013jya} found the
constraint $101<L<131$~MeV when all the observed frequencies are
interpreted as torsional oscillations, or $58<L<85$~MeV when the
second lowest frequency is assumed to have a different origin.  The
inclusion of electron screening effects can modify the former
constraint on $L$ to the range $97<L<127$~MeV \citep{Sotani:2015laa}.
Furthermore, in some cases having information on the mass and radius
of low-mass NSs may allow us to constrain EOS parameters. For example,
the observation of the X-ray burster 4U 1724-307 by
\cite{Suleimanov:2010th} allowed \cite{Sotani:2015lya} to set the
constraint $\eta \gtrsim 130$ MeV.
These astrophysical constraints seem incompatible with constraints on
$L$ obtained from the terrestrial nuclear experiments quoted
above~\citep{Li:2013ola,Lattimer:2014sga,Newton:2015xza}. One possible
reason for this discrepancy may be that the constraints from nuclear
experiments were obtained from almost stable nuclei, whose neutron
excess $\alpha$ in Eq.~(\ref{eq:snm}) is very small, while NS matter
deviates significantly from symmetric nuclear matter. Indeed, some
nuclear experiments with unstable nuclei suggest the possibility that
$L$ may have larger values~\citep{Tsang:2008fd,Yasuda:cea}.

If we adopt as fiducial values $40$~MeV$<L<80$~MeV -- recalling that
higher values cannot be excluded \citep{Newton:2015xza} -- and
$K_0 = 230 \pm 40$~MeV \citep{Khan:2013mga}, respectively, we can
conclude that a plausible range for $\eta$ is $67<\eta<120$, and that
higher values of $\eta$ may be possible. This should be kept in mind
in Section~\ref{sec:NSs} below, where we discuss how the bulk
properties of NSs depend on $\eta$.

\section{Neutron star structure}
\label{sec:NSs}

\subsection{Nonrotating neutron stars}
For nonrotating NSs, the effect of the nuclear symmetry parameters
introduced above on the mass-radius relation $M(R)$ was investigated
by~\cite{Sotani:2013dga}. The main finding of their work was that the
NS mass $M$ and the surface gravitational redshift $z$, defined as
\be
\label{redshift}
z=\left(1-\f{2M}{R}\right)^{-1/2}-1\,,
\ee
can be expressed as smooth functions of $\eta$ of the form
\begin{equation}
y = c_0 + c_1 \left(\frac{\eta}{100 {\rm \,MeV}}\right)\,,
\label{eq:fit_1}
\end{equation}
where $y$ collectively denotes either $M$ or $z$. We also note that
these relations can be combined to write the radius $R$ as a function
of $\eta$.

\begin{table*}
  \caption{Numerical values of the constant in the fitting expressions and
    of the rms percentage error $\sigma$ (last column), computed with / without EOS BSk21.}
\label{tab:fitcoef}
\begin{tabular}{l c c c c c c c c}

\hline
\hline
\multicolumn{8}{ |c| }{Using Eqs.~(\ref{eq:fit_1}) and (\ref{eq:cnormal})
-- Quantities at order
${\cal O}(\epsilon^0)$ in rotation} \\
\hline
Quantity & $c_{0,0}$ & $c_{0,1}$ & $c_{0,2}$ & $c_{1,0}$ & $c_{1,1}$ & $c_{1,2}$ & $\sigma$ \\ \\
$M$ & 0.34626 & -0.82183 & 0.29265 & -0.60780 & 1.2996 & -0.25890 & 0.086 / 0.076 \\
$z=\left(1-2M/R\right)^{-1/2}-1$ & 0.0040470 & -0.059527 & 0.026591 & -0.042719 & 0.10673 & -0.014208 & 0.100 / 0.089 \\
\hline
\multicolumn{8}{ |c| }{Using Eqs.~(\ref{eq:fit_2}) and (\ref{eq:cnormal})
-- Quantities at order
${\cal O}(\epsilon^1)$ and ${\cal O}(\epsilon^2)$ in rotation} \\
\hline
Quantity & $c_{0,0}$ & $c_{0,1}$ & $c_{0,2}$ & $c_{1,0}$ & $c_{1,1}$ & $c_{1,2}$ & $\sigma$ \\ \\
$\bar{I} = I/M^3$ & 4.1429 & -2.2458 & 0.46120 & -0.23654 & -0.26292 & 0.083322 & 0.122 / 0.083 \\
$\bar{Q} = -Q^{\rm (rot) \ast}/(M^3 \chi^2)$ & 2.9160 & -1.3835 & 0.24677 & -0.25594 & -0.093784 & 0.015956 & 0.120 / 0.114 \\
$\bar{\lambda}^{\rm (rot)} = \lambda^{\rm (rot)}/M^5$ & 11.203 & -5.8769 & 1.1697 & -0.24302 & -0.21457 & 0.055320 & 0.424 / 0.237 \\
$k_2^{\rm (rot)} = (3/2)\lambda^{\rm (rot)}/R^5$ & -3.9878 & 3.3914 & -0.91026 & -3.4378 & 2.6267 & -0.53179 & 0.273 / 0.272 \\
$e^{\ast}_{\rm Q} = -Q^{\rm (rot) \ast}/I$ & -2.1203 & 1.8784 & -0.52335 & -4.0307 & 3.0883 & -0.55984 & 0.129 / 0.126 \\
\hline
\multicolumn{8}{ |c| }{Using Eqs.~(\ref{eq:fit_2}) and (\ref{eq:cnormal})
-- Tidally deformed star} \\
\hline
Quantity & $c_{0,0}$ & $c_{0,1}$ & $c_{0,2}$ &$c_{1,0}$ & $c_{1,1}$ & $c_{1,2}$ & $\sigma$ \\ \\
$\bar{\lambda}^{\rm (tid)} = \lambda^{\rm (tid)}/M^5$ & 11.238 & -5.9413 & 1.1450 & -0.21434 & -0.25432 & 0.052281 & 0.462/ 0.263 \\
\hline
\hline
\end{tabular}
\end{table*}

The coefficients $c_i$ depend on the ratio
$u_{\rm c} \equiv \rho_{\rm c} / \rho_0$ which specifies the central
density of the stellar model. Following \cite{Sotani:2013dga}, we will
fit these coefficient using a quadratic polynomial in $u_{\rm c}$:
\begin{equation}
c_{i} = c_{i,0} + c_{i,1}\,u_{\rm c} + c_{i,2}\,u_{\rm c}^2\,.
\label{eq:cnormal}
\end{equation}
Therefore each of our empirical formulas will depend on six constant
parameters $c_{i,j}$.

In Fig.~\ref{fig:fits_M} we confirm the main results
of~\cite{Sotani:2013dga}. The left (right) panel shows that, quite
independently of the chosen EOS, the mass $M$ (the redshift $z$,
respectively) is indeed well fitted by a linear function of $\eta$ for
any fixed value of the central density $u_{\rm c}$: the plots show
this explicitly in the three cases $u_{\rm c}=1$, $u_{\rm c}=1.5$ and
$u_{\rm c}=2$. The bottom insets show that the fractional differences
$\delta y \equiv |y_{\rm data} - y_{\rm fit}|/y_{\rm data}$ for $M$
and $z$ are typically below $\sim 10\%$
(with the exception of EOS BSk21) whenever $\eta\gtrsim 67$~MeV.

The values of the fitting constants are listed in the top two rows of
Table~\ref{tab:fitcoef}.
To quantify the accuracy of these fits, Table~\ref{tab:fitcoef} also
lists the rms relative error
\be
\sigma \equiv \sqrt{\f{1}{N}\sum_{i=1}^{N}
  \left(1 - \f{y_{i}^{\rm fit}}{y_{i}^{\rm data}}\right)^2}\,,
\ee
where the sum runs over all stellar models $i=1,\,\dots,\,N$.
Two EOS models, namely BSk20 and
BSk21 (and particularly the latter), deviate more from our best-fit
function as $\rho_c$ increases. As pointed out by
\cite{Sotani:2013dga}, these deviations are of the order of the
uncertanties on the mass $M$ due to three-neutron interactions
obtained from the quantum Monte Carlo evaluations
\citep{Gandolfi:2011xu}. Therefore in Table~\ref{tab:fitcoef} we list
the values of $\sigma$ obtained either including or
omitting EOS BSk21, the EOS for which the errors are larger.

\subsection{Slowly rotating and tidally deformed neutron stars}
\label{sec:rot}

In general, rotating stellar models in general relativity must be
constructed numerically by solving a complicated system of partial
differential equations.
These numerical calculations \citep[reviewed
in][]{Stergioulas:2003yp,FriedmanStergioulas} suggest that uniformly
rotating NSs with physically motivated EOSs have dimensionless angular
momentum $\chi\lesssim 0.7$
\citep{Cook:1993qr,Berti:2003nb,Lo:2010bj}, but the spin magnitudes of
NSs in binary systems observable by Advanced LIGO are likely to be
much smaller than this theoretical upper
bound~\citep{Mandel:2009nx,Brown:2012qf}. The spin period of isolated
NSs at birth should be in the range 10-140~ms \citep[or
$\chi\lesssim 0.04$,][]{Lorimer:2001vd}. Accretion from a binary
companion can spin up NSs, but it is unlikely to produce periods less
than 1~ms \citep[i.e. $\chi\lesssim 0.4$,][]{Chakrabarty:2008gz}.  The
fastest spinning observed pulsar, PSR J1748-2446ad, has a period of
1.4~ms \citep[$\chi\sim 0.3$,][]{Hessels:2006ze}; the fastest known
pulsar in a NS-NS system, J0737-3039A, has a period of 22.70~ms
\citep[$\chi\sim 0.02$,][]{Burgay:2003jj}.

The perturbative formalism to construct slowly rotating NS models
was developed in the seminal works by \cite{Hartle:1967he}
and~\cite{Hartle:1968ht}. The formalism basically consists of an expansion
in terms of the small parameter $\epsilon \equiv \Omega/\Omega^{\ast}\ll 1$,
where $\Omega$ is the stellar angular velocity and
$\Omega^{\ast} \equiv \sqrt{M/R^3}$ is a characteristic rotation
frequency, comparable in order of magnitude to the mass-shedding
frequency of the star.
Subsequent work extended the formalism up to fourth order in
$\epsilon$, showing that the equilibrium properties of slowly rotating
solutions compare favorably with numerical codes for arbitrary
rotation rates~\citep{Berti:2004ny,Benhar:2005gi,Yagi:2014bxa}
even for the fastest known milisecond pulsar PSR
J1748-2446ad~\citep{Hessels:2006ze}. This pulsar spins
well below the estimated $\epsilon\approx 0.5$ for which the
Hartle-Thorne approximative scheme agrees very well with full
numerical calculations.

The slow-rotation approximation is basically an expansion in terms of
the small parameter $\epsilon \equiv \Omega/\Omega^{\ast}\ll 1$, where
$\Omega$ is the stellar angular velocity and
$\Omega^{\ast} \equiv \sqrt{M/R^3}$ is a characteristic rotation
frequency, comparable in order of magnitude to the mass-shedding
frequency of the star.
Even the fastest known millisecond pulsar \citep{Hessels:2006ze} spins
well below the estimated $\epsilon\approx 0.5$ for which the
Hartle-Thorne approximative scheme agrees very well with full
numerical calculations \citep{Berti:2004ny}.
Therefore the slow-rotation approximation is more than adequate to
extend the work on low-mass NSs by~\cite{Sotani:2013dga}.

The formalism to construct slowly rotating NS models was developed in
the seminal works by \cite{Hartle:1967he} and~\cite{Hartle:1968ht}.
Subsequent work extended the formalism up to fourth order in
$\epsilon$, showing that the equilibrium properties of slowly rotating
solutions compare favorably with numerical codes for arbitrary
rotation rates even for PSR
J1748-2446ad~\citep{Berti:2004ny,Benhar:2005gi,Yagi:2014bxa}.

We use the stellar structure equations as presented by
\cite{1999AAS..134...39S}, correcting the misprints listed by
\cite{Berti:2004ny}.  Our numerical results were validated by
comparison against the tables by \cite{Berti:2004ny}. For the
dimensionless bulk properties we follow the definitions of
\cite{Yagi:2013awa}. The explicit form of the structure equations,
their derivation and details of the integration procedure can be found
in these references.

\begin{figure*}
\includegraphics[width=2.1\columnwidth]{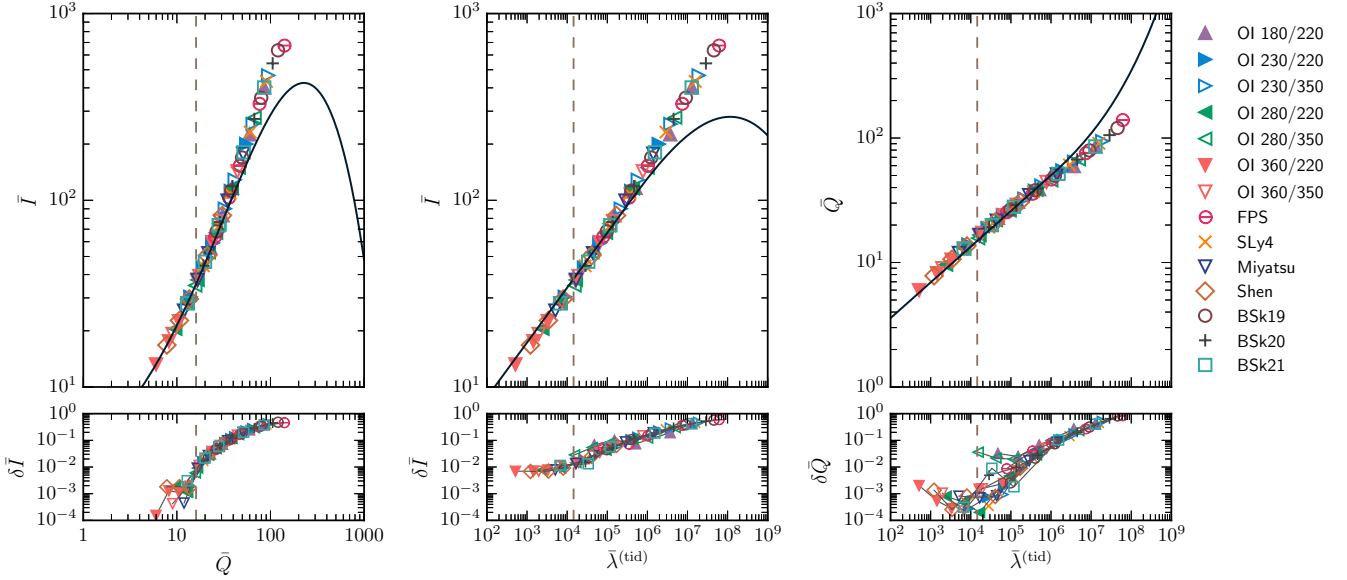}
\caption{The EOS-independent $I$-Love-$Q$ relations
  \citep{Yagi:2013bca,Yagi:2013awa} in the low-mass regime. The
  different panels show the $I$-$Q$ ({\it left}), $I$-Love ({\it
    center}) and Love-$Q$ ({\it right}) relations within the range of
  central energy densities considered here. For reference, the
  vertical dashed line corresponds to the values of $\bar{Q}$ and
  $\bar{\lambda}^{\rm (tid)}$ of a NS model using the Shen EOS with
  $M/M_{\odot} = 0.89$. The lower panels show that the fractional
    deviations in the $I$-Love-$Q$ relations increase for very low mass
  (i.e., larger values of $\bar{Q}$ and $\bar{\lambda}^{\rm (tid)}$).
  Nevertheless, near and above the minimum mass value
  $0.89 \, M_{\odot}$ the relations hold within an accuracy $< 2 \% $.
  The explicit functional form of the $I$-Love-$Q$ relations can be
  found in \citet[Eq.~(54) and Table 1]{Yagi:2013awa}. Observe that
    even for very low-mass NSs the universality remains, although it is
    not captured by the $I$-Love-$Q$ relations.}
 \label{fig:ILQ}
\end{figure*}

\begin{figure}
\includegraphics[width=\columnwidth]{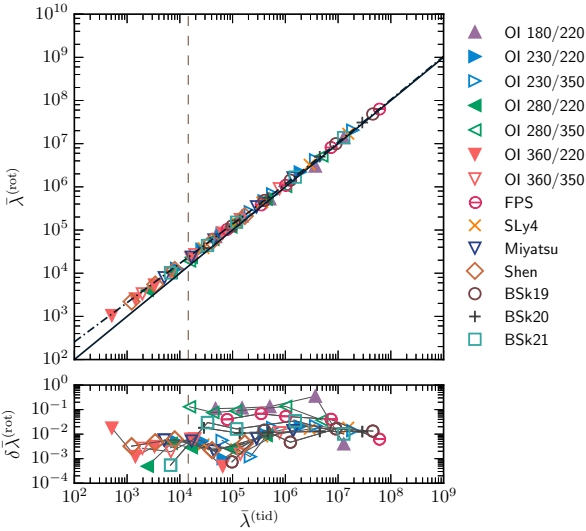}
\caption{The EOS independent Love-Love relations in the low-mass
  regime. The Love-Love relation between $\bar{\lambda}^{\rm (tid)}$
  and $\bar{\lambda}^{\rm (rot)}$ becomes an equality in the Newtonian
  -- i.e., small-$M$ -- limit \citep{Mora:2003wt}, and deviates from
      unity (solid line) for more relativistic stars. To make our fits of
  $\bar{\lambda}^{\rm (tid)}$ and $\bar{\lambda}^{\rm (rot)}$ with
  respect to $\eta$ useful in combination with the Love-Love relation,
  we derived the improved fit of Eq.~\eqref{eq:iloveq},
  corresponding to the dash-dotted line. This fit is accurate within
  $< 1\%$ for $M/M_{\odot} \geq 0.89$ (lower panel).  The vertical
  dashed line corresponds to NS model with mass $M/M_{\odot} = 0.89$
  using the Shen EOS.}
\label{fig:LL}
\end{figure}

At order ${\cal O}(\epsilon^0)$ in the perturbative expansion, a
static nonrotating star is characterized by its gravitational mass $M$
and radius $R$. Sometimes it is useful to replace the radius by the
surface redshift $z$ defined in Eq.~(\ref{redshift}).

At first order in rotation, i.e. ${\cal O}$($\epsilon^1$), the star is
also characterized by its moment of inertia $I$. Given $I$, we can
define a dimensionless moment of inertia $\bar{I} \equiv I/M^3$ as
well as a spin parameter $\chi \equiv I \Omega/M^2$
\citep{Yagi:2013awa}.

At second order in rotation, i.e. ${\cal O}$($\epsilon^2$), the star
deviates from its spherical shape and acquires a rotational quadrupole
moment $Q^{(\rm rot)\ast}$. For convenience, we define the dimensionless
rotation-induced quadrupole moment
$\bar{Q} \equiv - {Q^{(\rm rot) \ast}}/{\left(M^3 \chi^2\right)}$.  The
$\ell = 2$ rotational Love number $\lambda^{({\rm rot})}$ can be
defined in terms of $Q^{(\rm rot) \ast}$ as
$\lambda^{({\rm rot})} \equiv - Q^{(\rm rot) \ast}/\Omega^{2}$, and it can be
made dimensionless by defining
$\bar{\lambda}^{({\rm rot})} \equiv \bar{I}^2 \bar{Q}$. A quantity
closely related with $\lambda^{({\rm rot})}$ is the $\ell = 2$ apsidal
constant, defined as
$k_2^{({\rm rot})} \equiv (3/2)\lambda^{({\rm rot})} / R^5$; note that
\cite{Berti:2007cd} used a different definition.
Following \cite{Colaiuda:2007br}, we can also define the quadrupolar
rotational ellipticity\footnote{This quantity is different from the
  surface ellipticity $e_{\rm s}\equiv r_{\rm e}/r_{\rm p} - 1$, where
  $r_{\rm e}$ and $r_{\rm p}$ are the equatorial and polar radii of
  the oblate rotating star, respectively. The surface ellipticity is
  related to the so-called ``eccentricity''
  $e\equiv \left[(r_{\rm e}/r_{\rm p})^2 - 1\right]^{1/2}$ \cite[see
  e.g.][]{Berti:2004ny} by $e^2=e_{\rm s}^2+2e_{\rm s}$, and it
  describes the geometry of the star.}
$e^{\ast}_{\rm Q} \equiv - Q^{(\rm rot)\ast}/I$.
Note that all of the barred quantities defined above (as well as
$k_2^{({\rm rot})}$) are independent of the actual value of the
rotation parameter $\epsilon$.

\subsection{Tidally deformed neutron stars}
\label{sec:tid}

We will also be interested in the tidal deformation of a NS due the
presence of an orbiting companion, e.g. in a binary system. The
response to tidal deformations is encoded in the so-called $\ell = 2$
tidal Love number $\lambda^{({\rm tid})}$
\citep{2008ApJ...677.1216H,2009ApJ...697..964H,Damour:2009vw,Binnington:2009bb,Vines:2011ud},
which is potentially measurable by advanced gravitational-wave
interferometers. This quantity is in general
spin-dependent~\citep{Landry:2015zfa,Pani:2015hfa,Pani:2015nua,Landry:2015snx}, but
for simplicity we will assume that the tidally deformed NS is
nonrotating. The tidal Love number can be put in a dimensionless form
by defining $\bar{\lambda}^{\rm (tid)} \equiv \lambda^{\rm (tid)}/M^5$.  We
calculated the tidal Love number using the structure equations as
presented by \cite{Postnikov:2010yn}, and validated our results by
comparison against \cite{Yagi:2013awa}.

\subsection{Empirical relations for slowly rotating and
tidally deformed neutron stars}
\label{sec:fit}

We constructed NS models for all of the 14 EOS models listed in
Table~\ref{tab:eosparam}. We integrated the structure equations for
central total energy densities within the range
$u_{\rm c}\equiv \rho_{\rm c} / \rho_0 \in [1.0, 2.0]$ in increments
$\Delta u_{\rm c} = 0.1$, for a total of 154 stellar models.
We verified that the normalized binding energy $M_{\rm b} / M - 1$
(where $M_{\rm b}$ is the baryonic mass) is positive, so that all of
these stellar configurations are bound.

Our results for the $I$-Love-$Q$ relations are shown in
Fig.~\ref{fig:ILQ}, which confirms the main findings of
\cite{Yagi:2013awa}: the universality holds within a few percent,
except for very low-mass stars.
This breakdown of the $I$-Love-$Q$ relations was already visible
e.g. in Fig.~9 of \cite{Yagi:2013awa}, but it is much more noticeable
in the low-mass range explored in this work.  \cite{Yagi:2014qua}
suggested that the $I$-Love-$Q$ relations hold because of an
approximate self-similarity in the star's isodensity contours. This
approximate symmetry only holds for compact stars, but it is broken in
low-mass NSs, white dwarfs and ordinary stars.  Indeed, the
$I$-Love-$Q$ relations presented in \cite{Yagi:2014qua} were obtained
by fitting data in the range $\bar{Q} < 20$ and
$\bar{\lambda}^{\rm (tid)} < 2 \times 10^4$ \citep{kent:private}.

In Fig.~\ref{fig:LL} we show that a universal ``Love-Love'' relation
also holds between the tidal and rotational Love numbers. A
well-known result in Newtonian gravity is that tidal and rotational
Love numbers are the same \citep{Mora:2003wt}. This equality no longer
holds true for relativistic stars
\citep{Berti:2007cd,Yagi:2013awa}. Therefore we propose a different fit,
namely:
\be
\ln{\bar{\lambda}^{({\rm rot})}} = \sum_{j = 0}^{4}\,
k_{j} \left(\ln \bar{\lambda}^{({\rm tid})}\right)^{j},
\label{eq:iloveq}
\ee
where $k_0 = 2.1089$, $k_1 = 6.5084 \times 10^{-1}$,
$k_2 = 2.4688 \times 10^{-2}$, $k_3 = - 8.9891 \times 10^{-4}$ and
$k_4 = 1.3985 \times 10^{-5}$. This fit uses data in the central
density range $u_{\rm c}\in[0.9,\,2.0]$, and it works accurately in
the range of masses of our interest.

Using these numerical calculations we then fitted the various bulk
properties of NSs as functions of $\eta$ and of the central density.
The quantities characterizing rotating stars -- namely
$\bar{I}$, $\bar{Q}$, the $\ell = 2$ rotational Love number
$\bar{\lambda}^{({\rm rot})} \equiv \bar{I}^2 \bar{Q}$, the
$\ell = 2$ apsidal constant $k_2^{({\rm rot})}$ and the quadrupolar
ellipticity $e^{\ast}_{Q}$ -- are well fitted by
functions of the form
\begin{equation}
\log_{10} y = c_0 \left(\frac{\eta}{100 {\rm \,MeV}}\right)^{c_1}\, .
\label{eq:fit_2}
\end{equation}
Just as for the nonrotating bulk properties of NSs, the coefficients
$c_i$ depend on the ratio $u_{\rm c} \equiv \rho_{\rm c} / \rho_0$,
which specifies the central density of the stellar model. Following
\cite{Sotani:2013dga}, we fitted the $c_i$'s by quadratic polynomials
of the form (\ref{eq:cnormal}), so that each of our empirical formulas
depends on six constant parameters $c_{i,j}$.

\begin{figure}
\includegraphics[width=\columnwidth]{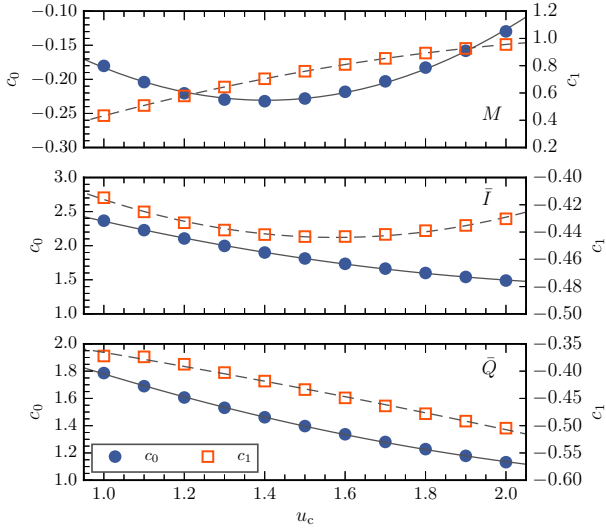}
\caption{Illustration of the behavior of $c_0$ (circles) and $c_1$
  (squares), appearing in the fitting expressions (\ref{eq:fit_1}) and
  (\ref{eq:fit_2}), as functions of $u_c$. The $c_i$'s are shown for
  three representative bulk properties of NSs: the mass $M$ ({\it
    top}) fitted using Eqs.~(\ref{eq:fit_1}) and (\ref{eq:cnormal});
  the dimensionless moment of inertia $\bar{I}$ ({\it center}) fitted
  using Eqs.~(\ref{eq:fit_2}) and (\ref{eq:cnormal}); and the
  dimensionless quadrupole moment $\bar{Q}$ ({\it bottom}) fitted
  using Eqs.~(\ref{eq:fit_2}) and (\ref{eq:cnormal}).}
\label{fig:coeffs}
\end{figure}

The quality of the fits is shown in the three panels of
Fig.~\ref{fig:coeffs} for three representative bulk NS properties,
namely the mass $M$, the dimensionless moment of inertia $\bar I$ and
the dimensionless quadrupole moment $\bar Q$.  In summary: the
nonrotating bulk properties of NSs can be expressed as functions of
$\eta$ of the form (\ref{eq:fit_1}) combined with
Eq.~(\ref{eq:cnormal}); the bulk properties of rotating and tidally
deformed NSs can be fitted by functions of the form (\ref{eq:fit_2})
combined with Eq.~(\ref{eq:cnormal}).
The fitting coefficients are listed in Table~\ref{tab:fitcoef}, that
collects the main results of our investigation.

\begin{figure*}
 \includegraphics[width=\columnwidth]{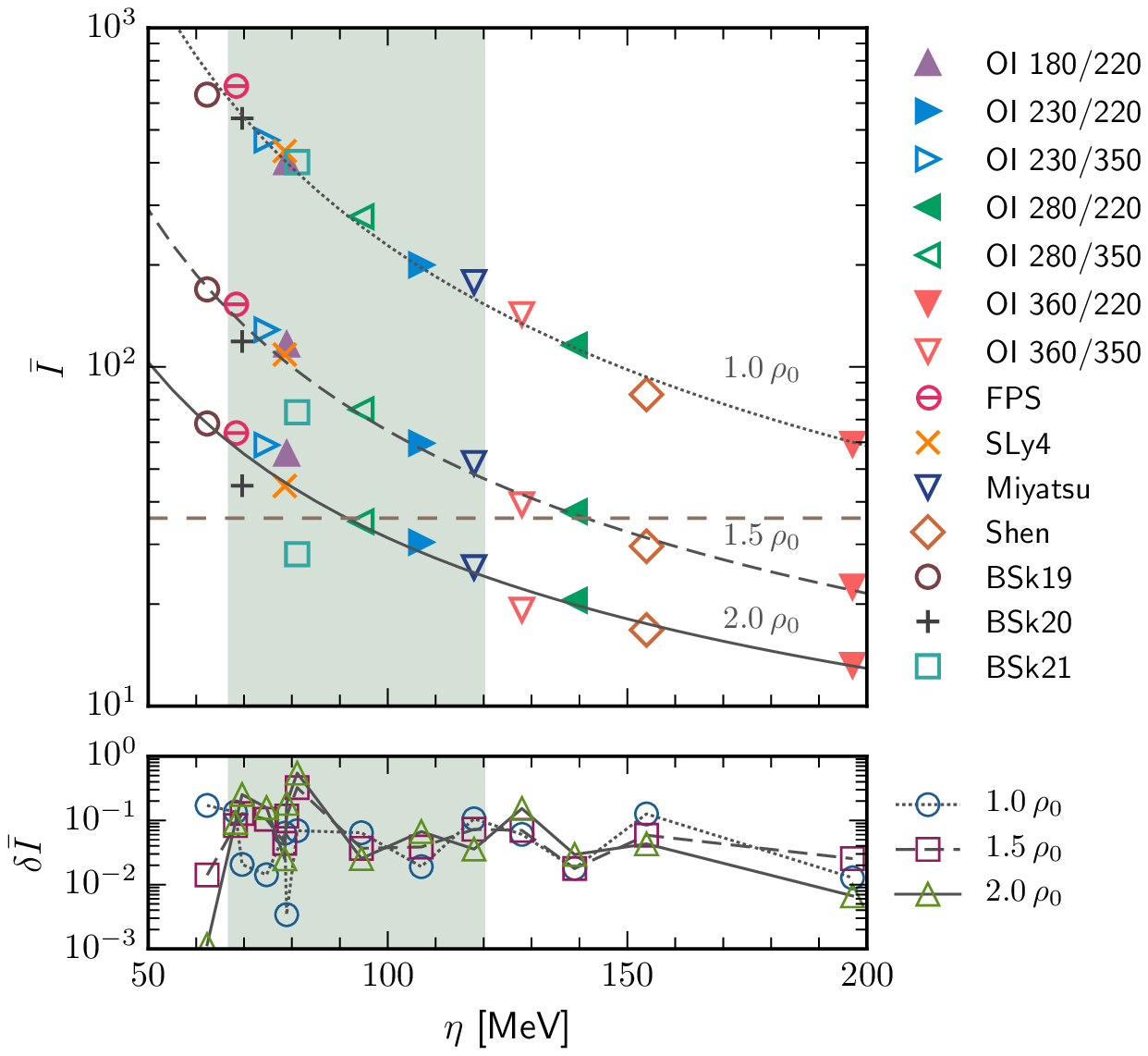}
 \includegraphics[width=\columnwidth]{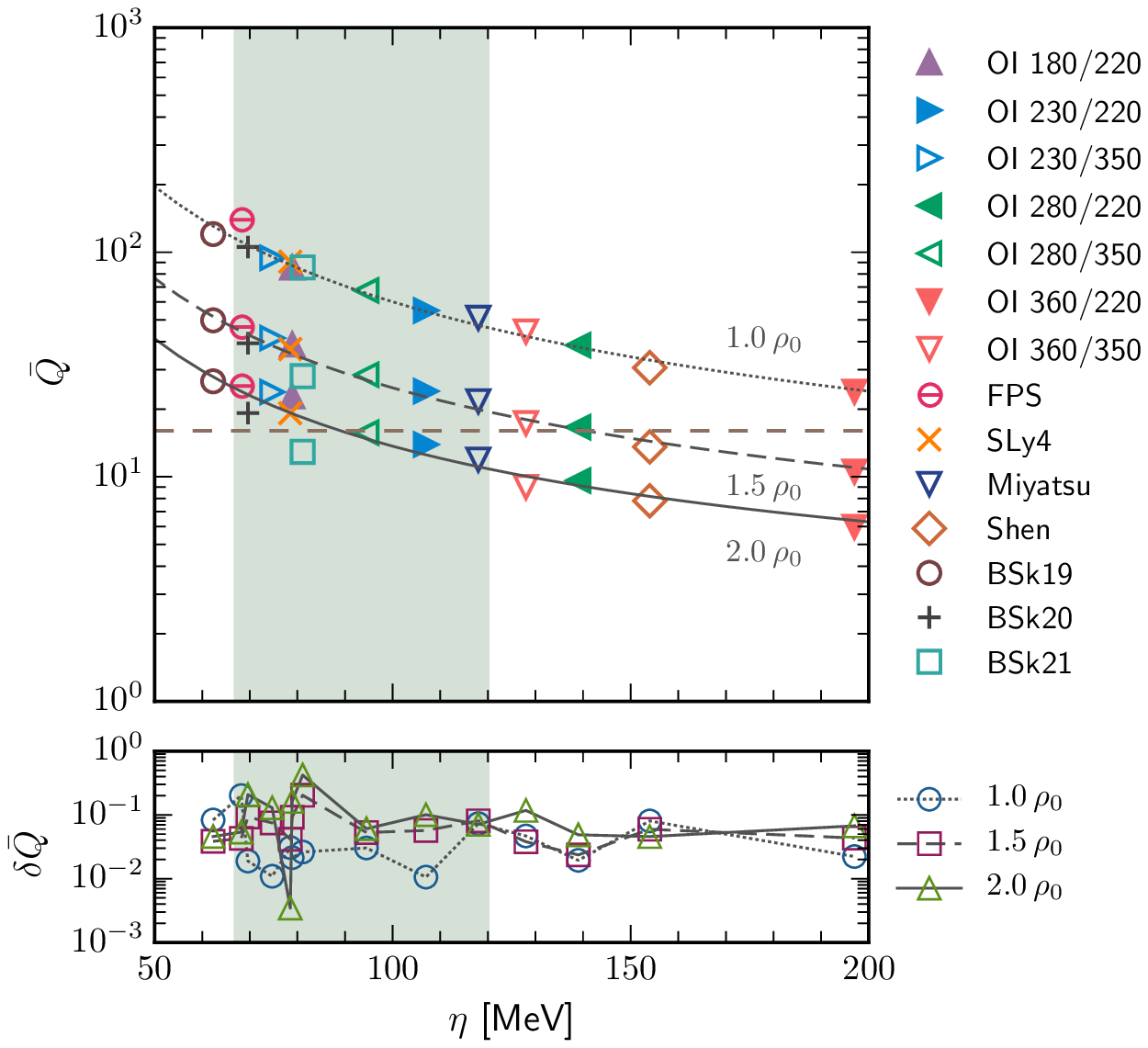}
 \caption{{\it Left:} fit of $\bar{I} \equiv I/M^3$.  {\it Right:} Fit
   of the reduced quadrupole moment
   $\bar{Q} \equiv Q^{\rm (rot) \ast}/(M^3 \chi^2)$. The horizontal
   dashed line in the left (right) panel marks the value of $\bar{I}$
   ($\bar{Q}$) for a NS with $M/M_{\odot} = 0.89$ and the Shen EOS.}
\label{fig:fits_ql}
\end{figure*}

\begin{figure*}
 \includegraphics[width=\columnwidth]{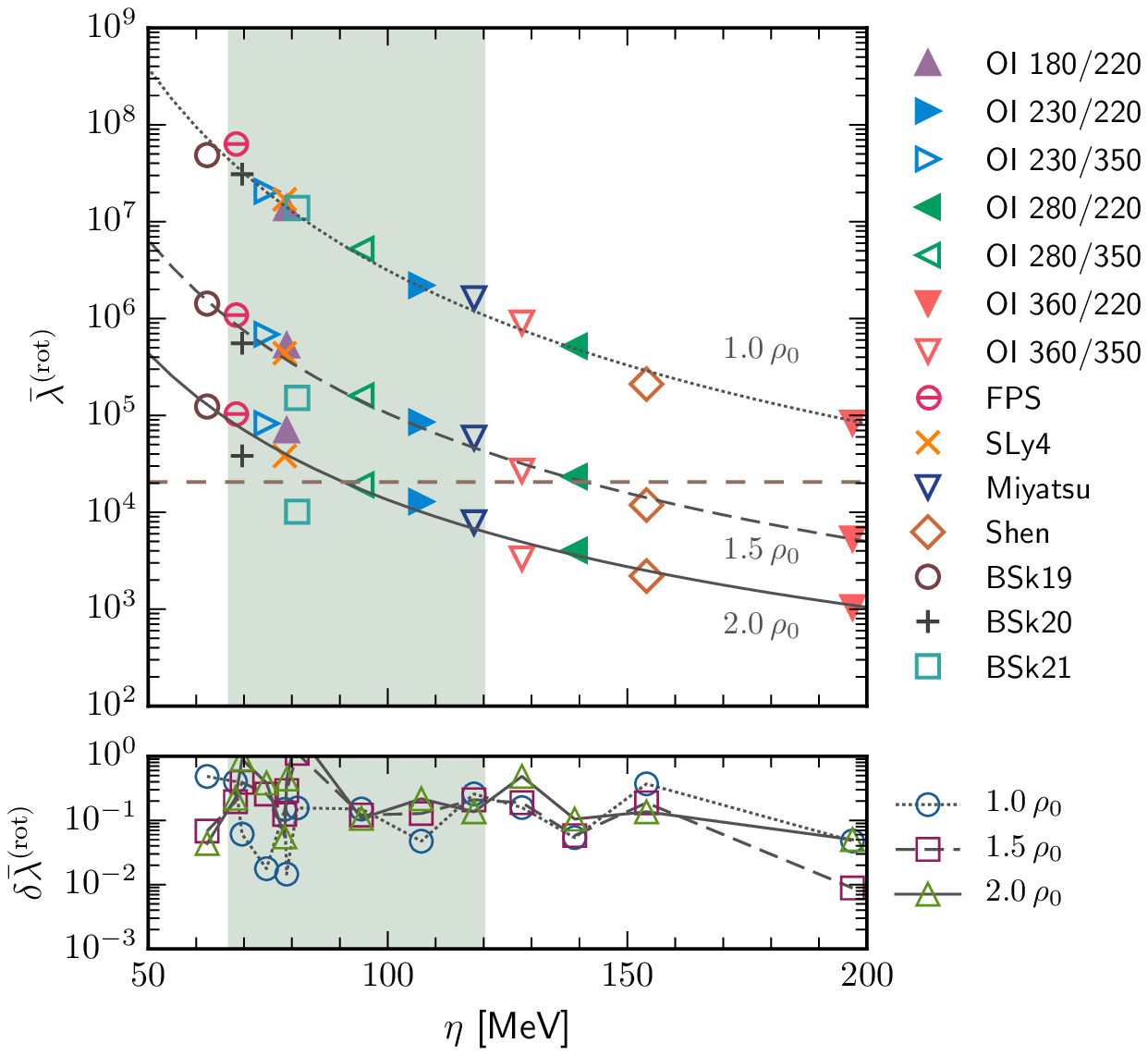}
 \includegraphics[width=\columnwidth]{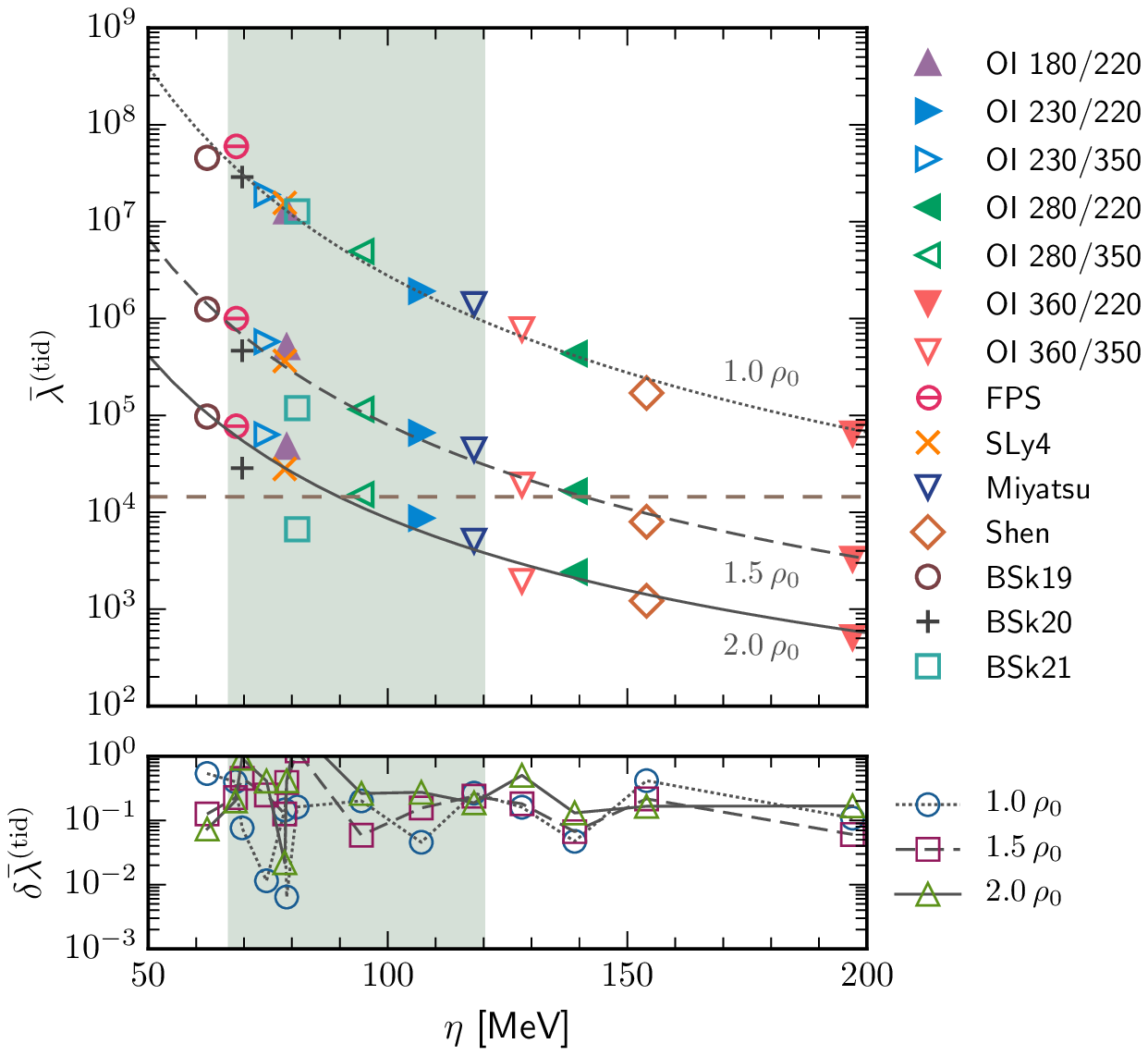}
 \caption{ {\it Left:} Fit of the rotational Love number
   $\bar{\lambda}^{(\rm rot)}$.  {\it Right:} Fit of the tidal Love
   number $\bar{\lambda}_{\rm tid}$.  The horizontal dashed line in
   the left (right) panel marks the value of
   $\bar{\lambda}^{(\rm rot)}$ ($\bar{\lambda}^{(\rm tid)}$) for a NS
   with $M/M_{\odot} = 0.89$ and the Shen EOS. }
 \label{fig:fits_ek}
\end{figure*}

\begin{figure*}
\includegraphics[width=\columnwidth]{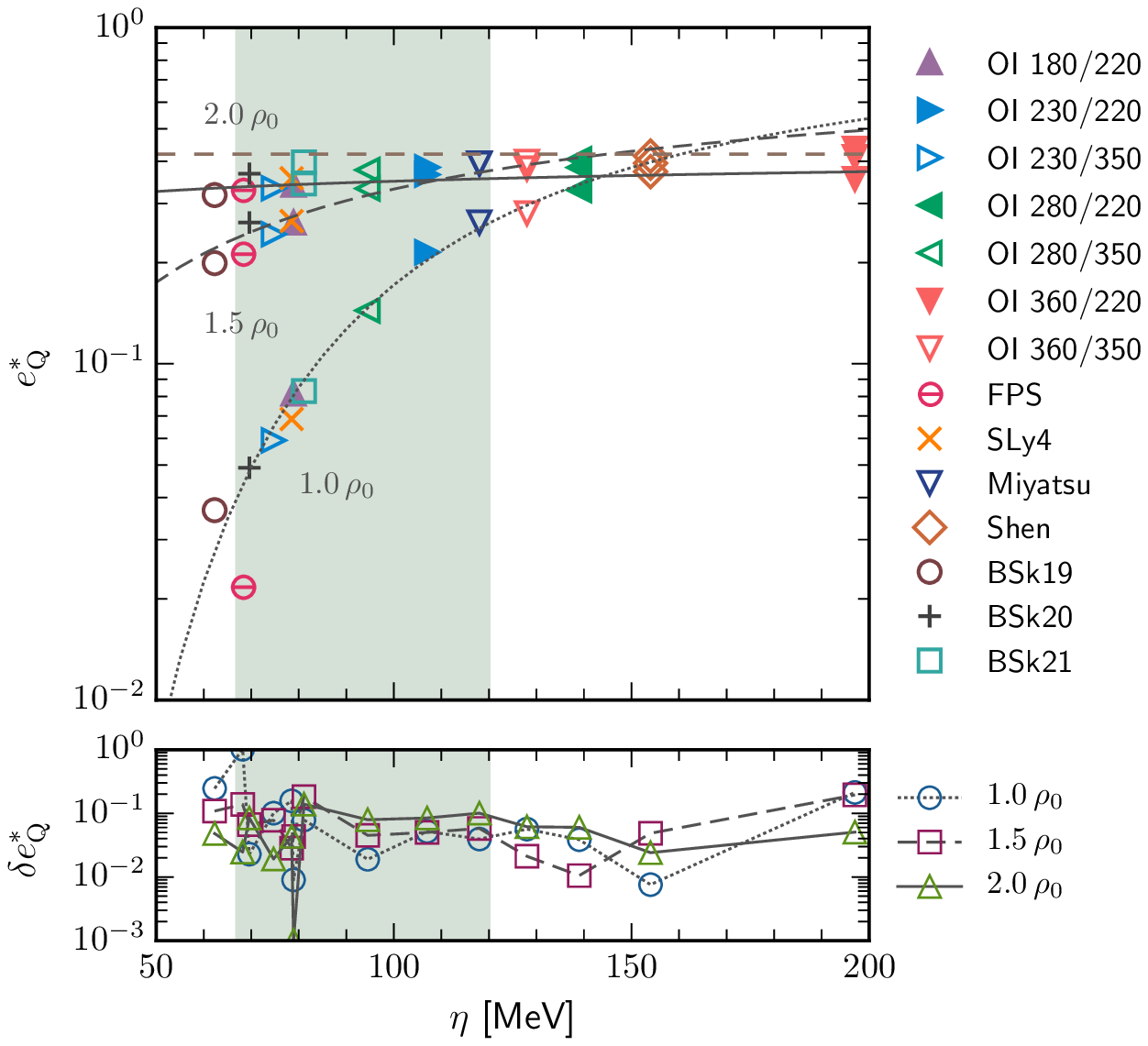}
\includegraphics[width=\columnwidth]{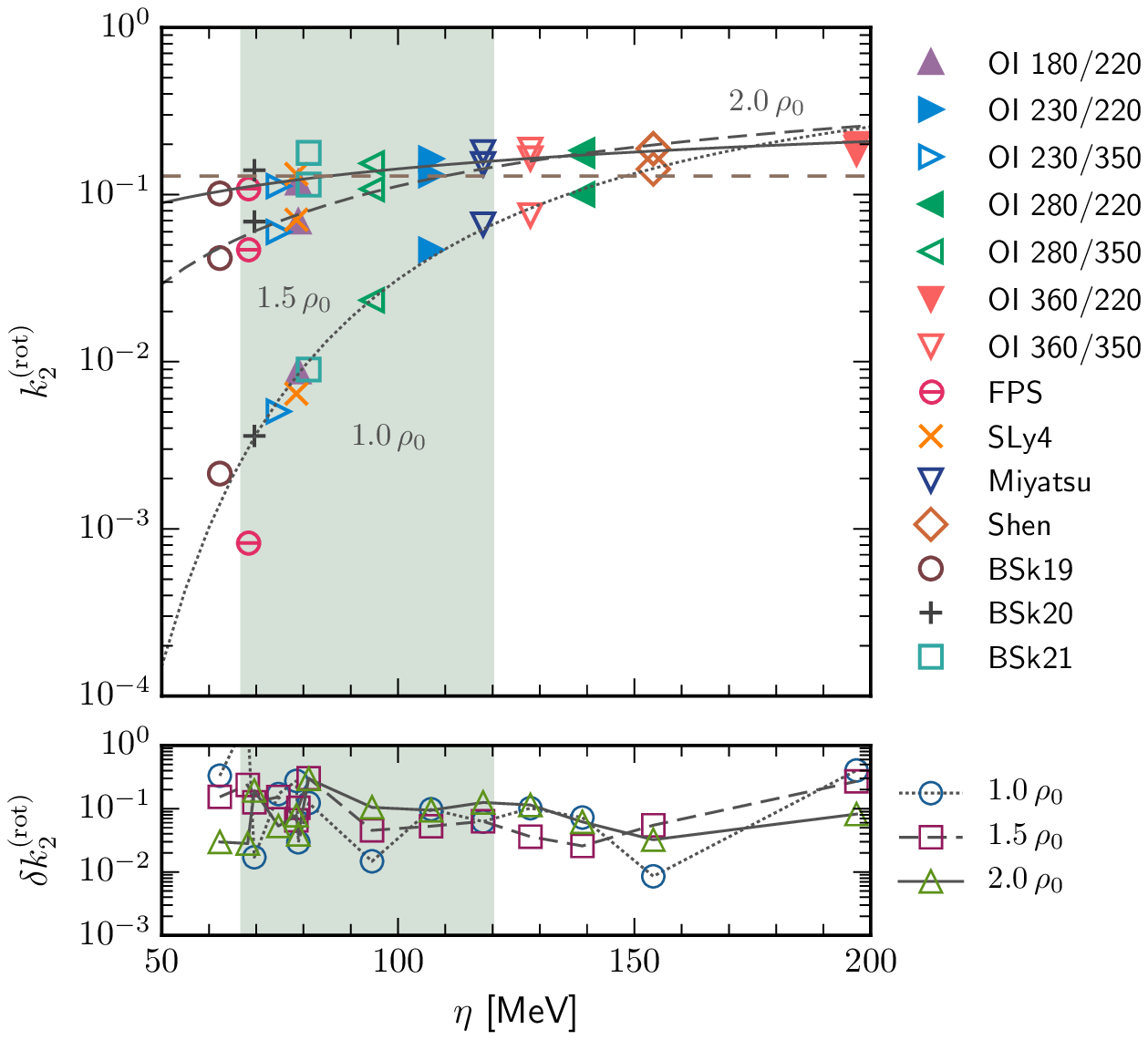}
\caption{{\it Left:} Fit of the quadrupole ellipticity $e^{\ast}_{\rm Q}$.
{\it Right:} fit of the $\ell = 2$ rotational apsidal constant $k^{\rm (rot)}_{2}$.
Both quantities behave similarly, becoming nearly independent of
$\eta$ for $\rho_c = 2.0 \rho_0$. For large value of $\eta$, we see
that $k_{2}^{\rm rot} \approx 0.7$ irrespective of the central
density.  The horizontal dashed line in the left (right) panel marks
the value of $e^{\ast}_{\rm Q}$ ($k^{\rm (rot)}_{2}$) for a NS with
$M/M_{\odot} = 0.89$ and the Shen EOS.  }
\label{fig:fits_eQes}
\end{figure*}

The fits and their accuracy are also presented graphically in
Figures~\ref{fig:fits_ql}--\ref{fig:fits_eQes}. In the lower panel of
each figure we plot the fractional differences $\delta y$.
As in Fig.~\ref{fig:fits_M}, the shaded area corresponds to the most
plausible range of values for $\eta$ as discussed in
Sec.~\ref{sec:expcons}.
These figures show that in general the fits work quite well in our
fiducial range, i.e. for $\eta>67$~MeV, with larger errors for small
$\eta$.

\section{Conclusions and outlook}
\label{sec:conclusions}

In this paper we have integrated the Hartle-Thorne equations for an
extensive set of EOS models. We have computed the bulk properties of
nonrotating (mass $M$ and radius $R$, or equivalently mass $M$ and
surface redshift $z$), rotating (moment of inertia $I$, quadrupole
moment $Q$, quadrupole ellipticity $e_{\rm Q}$, rotational Love number
$\lambda^{\rm (rot)}$, $\ell=2$ apsidal constant $k_2^{\rm (rot)}$)
and tidally deformed ($\ell=2$ tidal Love number
$\lambda^{\rm (tid)}$) low-mass NSs. All of these bulk NS properties
can be fitted by relatively simple functions of the central density
(more precisely, of $u_{\rm c}\equiv \rho_{\rm c}/\rho_0$) and of the
parameter $\eta\equiv (K_0 L^2)^{1/3}$, where $K_0$ is the
incompressibility of symmetric nuclear matter and $L$ is the slope of
the symmetry energy at saturation density. The coefficients of these
fitting relations are summarized in Table~\ref{tab:fitcoef}.

The main conclusion of this work is that the measurement of {\it any
two} of these bulk properties in low-mass NSs can be used -- at
least in principle -- to infer the values of $\rho_{\rm c}$ and
$\eta$, providing important information on the EOS. However there are
some important practical caveats.

First and foremost -- as shown in the lower panels of
Figures~\ref{fig:fits_ql}--\ref{fig:fits_eQes} -- the fitting
relations are approximate, with relative errors that typically get
larger for the lowest plausible values of $\eta$. Furthermore, as
discussed in the literature, constraints on the bulk properties of NSs
and on the EOS require Monte Carlo simulations or dedicated Bayesian
studies which are beyond the scope of this paper
\citep{Steiner:2010fz,Steiner:2012xt,Steiner:2014pda,Lattimer:2014sga}.

Even assuming accurate measurements of two of the bulk properties of a
given NS -- to be concrete, say $M$ and $I$ -- a conceptual limitation
is that not all EOSs predict the existence of NSs with ``realistic''
masses (say, $M>0.89M_\odot$) in the range $1\leq u_{\rm c}\leq 2$
where our fitting relations have been derived
(cf. Fig.~\ref{fig:massradius}). This problem can in principle be
circumvented, because the measurement of $M$ and $I$ can be used to
infer {\it both} $u_{\rm c}$ and $\eta$, and thus to verify whether
the NS really has central densities in the range of interest. However
it is possible that systematic errors could spoil these consistency
tests. For example one could imagine a situation where the ``true''
central density corresponds to (for example) $u_{\rm c}=3$, but since
we are applying the fitting relations outside of their region of
validity, we recover values of $u_{\rm c}\in [1,\,2]$ and get a wrong
estimate for $\eta$ \citep{kent:private}. These data analysis issues
deserve further study.

The recent discovery of universal $I$-Love-$Q$ relations is also
helpful. For example we can imagine measuring (say) $M$, $R$ and
$\bar{I}$ and getting information on the remaining bulk properties by
exploiting the $I$-Love-$Q$ relations. A measurement of multiple
parameters for the same astrophysical NS can be combined with our
fitting formulas either to check the consistency of the inferred
values of $u_{\rm c}$ and $\eta$, or to reduce statistical and/or
systematic errors.

We conclude by speculating on some observational possibilities to
implement this program.

Perhaps the most promising avenue in the near future is the
measurement of mass and moment of inertia through relativistic
spin-orbit coupling in systems such as the ``double pulsar'' PSR
J0737-3039 \citep{Burgay:2003jj}, especially considering that a 10\%
measurement of the moment of inertia {\em alone} can yield tight
constraints on the pressure over a range of densities to within
$50-60\%$ \citep{Steiner:2014pda}. This possibility was discussed by
various authors
\citep{Damour:1988mr,Lattimer:2004nj,Bejger:2005jy}. The experimental
challenges associated with these measurements in the case of the
double pulsar are reviewed in (e.g.) in Section 6 of~\cite{Kramer:2009zza}.

In the near future it may also be possible to constrain $\eta$ by
gravitational-wave observations.
Low-mass isolated NSs are relatively promising gravitational-wave
sources because they are more deformable and their crusts can support
larger ellipticities, generating stronger gravitational-wave
signals~\citep[][cf. also our Figs.~\ref{fig:fits_ql} and~\ref{fig:fits_ek}]{Horowitz:2009vm,JohnsonMcDaniel:2012wg,Johnson-McDaniel:2013nql}.
However the characteristic gravitational-wave amplitude
depends on a (generally unknown) geometrical factor involving the
orientation of the NS and the antenna pattern of the detectors
\citep[see e.g.][]{Bonazzola:1995rb,Dhurandhar:2011qe}.
It has recently been proposed that gravitational-wave measurements of
a {\em stochastic} gravitational-wave background from rotating NSs
could be used to constrain the average NS ellipticity, and (if
constraints on the masses can be obtained) these measurements could
also constrain $\eta$. This possibility seems most promising for
third-generation detectors such as the Einstein Telescope
\citep{Talukder:2014eba}.

\cite{Ono:2015kza} proposed to estimate the mass of an isolated
rapidly rotating NS by exploiting the mass-dependent logarithmic phase
shift caused by the Shapiro time delay. According to their Monte Carlo
simulations, the mass of a NS with spin frequency $f=500$~Hz and
ellipticity $10^{-6}$ at 1 kpc is typically measurable with an
accuracy of 20\% using the Einstein Telescope. Higher-order terms in
the Shapiro time delay will depend on the higher multipole moments $I$
and $Q$, and they may allow us to measure these moments and constrain
$\eta$.
It may also be possible to combine the empirical relations of the
present work with similar fitting relations that have been developed
in the context of gravitational-wave asteroseismology \citep[see
e.g.][]{Doneva:2015jba}.

Last but not least, as mentioned in the introduction, Advanced LIGO
observations of binary systems involving NSs could yield measurements
of masses and tidal Love numbers
\citep{Mora:2003wt,Berti:2007cd,Flanagan:2007ix,Read:2009yp,Hinderer:2009ca,Vines:2011ud,Damour:2012yf,DelPozzo:2013ala,Read:2013zra,Lackey:2013axa,Yagi:2015pkc,Dietrich:2015pxa,Agathos:2015uaa}.
Recent studies pointed out that systematic errors on these
measurements are large and that better waveform models are necessary
\citep{Favata:2013rwa,Yagi:2013baa,Wade:2014vqa,Chatziioannou:2015uea}, but
effective-one-body methods and numerical simulations are making
remarkable progress in this direction
\citep{Bernuzzi:2012ci,Bernuzzi:2014kca,Bernuzzi:2014owa,Bernuzzi:2015rla}.

\section*{Acknowledgements}
We are indebted to Kent Yagi, Leo C. Stein and Nico Yunes for a very
careful reading of an early draft of this paper, and for valuable
comments that significantly improved the paper.
We thank Kei Iida, Kazuhiro Oyamatsu and Anthea F. Fantina for sharing
some of the EOS data tables used in this work.
We thank Caio F.B. Macedo and Kent Yagi for validating some
of our numerical results.
We also thank Michalis Agathos for very carefully reading our
manuscript and for making important suggestions and comments.
E.B. was supported by NSF CAREER Grant No.~PHY-1055103 and by FCT
contract IF/00797/2014/CP1214/CT0012 under the IF2014 Programme.
H.O.S was supported by NSF CAREER Grant No.~PHY-1055103 and by a
Summer Research Assistantship Award from the University of
Mississippi. E.B. and H.O.S. thank the Instituto Superior T\'ecnico
(Lisbon, Portugal), where this project started, for the
hospitality.
H.S. was supported by Grant-in-Aid for Young Scientists (B) through
No. 26800133 provided by JSPS and by Grants-in-Aid for Scientific
Research on Innovative Areas through No. 15H00843 provided by MEXT.

\bibliographystyle{mnras}

\bsp    
\label{lastpage}
\end{document}